\documentclass[amsmath,amssymb,reprint,prl,aps]{revtex4-1}
\usepackage{comment}
\usepackage{graphicx}
\usepackage{dcolumn}
\usepackage{bm}
\usepackage{hyperref}
\usepackage{graphicx} 
\usepackage{subfigure}
\usepackage{bm}
\usepackage{latexsym}
\usepackage{ulem}
\usepackage{datetime}
\usepackage{scrtime}
\usepackage[dvipsnames, svgnames, x11names]{xcolor}

\definecolor{darkgreen}{RGB}{50,150,0}
\definecolor{purple}{cmyk}{0.5,0.75,0,0}
\definecolor{darkpurple}{RGB}{128,0,128}
\definecolor{ultramarine}{rgb}{0.07, 0.04, 0.56}
\definecolor{cadmiumgreen}{rgb}{0.0, 0.42, 0.24}
\definecolor{indigo(dye)}{rgb}{0.0, 0.25, 0.42}

\hypersetup{colorlinks=true,linkcolor=blue,citecolor=blue}

\def\be{\begin{equation}}
\def\ee{\end{equation}}
\def\ba{\begin{eqnarray}}
\def\ea{\end{eqnarray}}

\def\lf{\left}
\def\rt{\right}

\begin{document}

\title{Primordial black holes from null energy condition violation during inflation}

\begin{abstract}
Primordial black holes (PBHs) and the violation of the null energy condition (NEC) have significant implications for our understanding of the very early universe. We present a novel approach to generate PBHs via the NEC violation in a single-field inflationary scenario. In our scenario, the universe transitions from a first slow-roll inflation stage with a Hubble parameter $H = H_{\text{inf}1}$ to a second slow-roll inflation stage with $H = H_{\text{inf}2}\gg H_{\text{inf}1}$, passing through an intermediate stage of NEC violation. The NEC violation naturally enhances the primordial scalar power spectrum at a certain wavelength, leading to the production of PBHs with masses and abundances of observational interest. We also investigate the phenomenological signatures of scalar-induced gravitational waves (SIGWs) resulting from the enhanced density perturbations. Our work highlights the potential of utilizing a combination of PBHs, SIGWs, and primordial gravitational waves as a valuable probe for studying NEC violation during inflation, opening up new avenues for exploring the early universe.
\end{abstract}

\author{Yong Cai$^{1}$}
\email[]{yongcai\_phy@outlook.com}
\email[]{caiyong@zzu.edu.cn}
\author{Mian Zhu$^{2}$}
\email[Corresponding author:~]{mian.zhu@uj.edu.pl}
\author{Yun-Song Piao$^{3,4,5,6}$}
\email[Corresponding author:~]{yspiao@ucas.ac.cn}
\affiliation{$^1$ Institute for Astrophysics, School of Physics, Zhengzhou University, Zhengzhou 450001, China}
\affiliation{$^2$ Faculty of Physics, Astronomy and Applied Computer Science, Jagiellonian University, 30-348 Krakow, Poland}
\affiliation{$^3$ School of Physical Sciences, University of Chinese Academy of Sciences, Beijing 100049, China}
\affiliation{$^4$
International Centre for Theoretical Physics Asia-Pacific,
University of Chinese Academy of Sciences, 100190 Beijing, China}
\affiliation{$^5$ School of Fundamental Physics and Mathematical Sciences,
Hangzhou Institute for Advanced Study, UCAS, Hangzhou 310024, China}
\affiliation{$^6$ Institute of Theoretical Physics, Chinese Academy of Sciences, P.O. Box 2735, Beijing 100190, China}

\maketitle


{\it Introduction.---} Primordial black holes (PBHs) are powerful probes for studying the physics of the early universe \cite{Hawking:1971ei,Carr:1974nx, Carr:1975qj}. In contrast to the astrophysical black holes, which evolve from massive stars and contain masses larger than 5 $M_{\odot}$ \cite{Ozel:2010su}, PBHs can have a wide mass range from tens of micrograms to millions of solar masses. In view of that, PBHs can be relevant to astrophysical and cosmological phenomena such as the origin of dark matter \cite{Ivanov:1994pa, Belotsky:2014kca,Carr:2016drx,Carr:2020xqk}, and the seeds of the supermassive black holes \cite{Bean:2002kx,Kawasaki:2012kn}. Therefore, the formation of PBHs in the early universe is vastly studied, see, e.g., \cite{Kawasaki:2012wr,Chen:2016kjx,Quintin:2016qro,Wang:2016ana,Wang:2019kaf,Garriga:2015fdk,Pi:2017gih,Nakama:2018utx,Ding:2019tjk,Luo:2020dlg,Domenech:2020ssp,Boudon:2020qpo,Kawana:2021tde,Meng:2022ixx,Lin:2021vwc,Wu:2021zta,Tan:2022lbm,Pi:2021dft,Domenech:2021wkk,Liu:2021svg,Wang:2022nml,He:2022amv,Papanikolaou:2022hkg,Kristiano:2022maq,Fumagalli:2023hpa,Riotto:2023gpm,DeLuca:2022uvz,DeLuca:2022bjs,DAgostino:2022ckg,Lewicki:2023ioy,Choudhury:2023rks,Choudhury:2023hvf,He:2023yvl,Zhang:2023tfv}.

In the literature, PBHs are thought to arise from over-dense regions collapsing due to self-gravity. Inflation, as the most widely accepted paradigm of the very early universe, is capable of generating primordial scalar (or density) perturbations that are consistent with observations of the cosmic microwave background (CMB). Therefore, the key to PBH formation in inflationary cosmology is to obtain a significant extra enhancement of the amplitudes of primordial scalar perturbations on small scales (see, e.g., \cite{Drees:2011yz,Inomata:2017okj,Kannike:2017bxn,Cheng:2018yyr,Cai:2019jah,Lin:2020goi,Ashoorioon:2020hln,Ozsoy:2020kat,Kawai:2021edk,Karam:2022nym,Papanikolaou:2022did,Fu:2022ssq,Choudhury:2023vuj}), while simultaneously satisfying the observational constraints on the CMB scale.

In the single-field slow-roll inflation scenario, the power spectrum of primordial scalar perturbations is dependent on the Hubble parameter $H$, the slow-roll parameter $\epsilon\equiv-\dot{H}/H^2$ (or its generalized formulation), and the sound speed $c_s$ of the scalar perturbation mode. In standard single-field slow-roll inflation, the scalar power spectrum is nearly scale-invariant on all scales.
To produce sizable PBHs in single-field inflation models, various mechanisms have been investigated, including (but not limited to) ultra-slow-roll {(USR)} inflation \cite{Garcia-Bellido:2017mdw,Germani:2017bcs,Byrnes:2018txb,Fu:2019ttf,Fu:2019vqc,Fu:2020lob,Ragavendra:2020sop,Di:2017ndc,Motohashi:2017kbs,Dalianis:2018frf,Yi:2020cut}, and modifications to the dispersion relation or the sound speed $c_s$ of scalar perturbations \cite{Ballesteros:2018wlw,Kamenshchik:2018sig,Ashoorioon:2019xqc,Qiu:2022klm} (see also \cite{Cai:2018tuh,Cai:2019bmk,Cai:2020ovp,Zhou:2020kkf,Peng:2021zon} for the mechanism of parametric resonance).
In addition to these models, many inflationary models exhibit sudden strong enhancement of the power spectrum on certain scales when slow-roll conditions are violated at some stages, such as the Starobinsky model when there is a non-smooth potential \cite{Starobinsky:1992ts,Polarski:1992dq}, which not only has the capacity to generate a significant amount of PBHs but may also produce a large GW background, see, e.g., \cite{Lesgourgues:1998mq,Tasinato:2020vdk}.

The violation of the null energy condition (NEC), or more precisely, the null congruence condition in modified gravity, is closely related to potential solutions for the singularity problem in the context of the Big Bang and inflationary cosmology \cite{Rubakov:2014jja}. It may play a crucial role in the very early universe. Fully stable NEC violation can be achieved in ``beyond Horndeski'' theories \cite{Cai:2016thi, Creminelli:2016zwa, Cai:2017tku, Cai:2017dyi, Kolevatov:2017voe, Ilyas:2020qja}. In this letter, we propose a new approach to generate PBHs in a single-field inflation scenario by enhancing the curvature perturbations through intermediate NEC violation.

\begin{figure}[htbp]
\includegraphics[scale=2,width=0.9\linewidth]{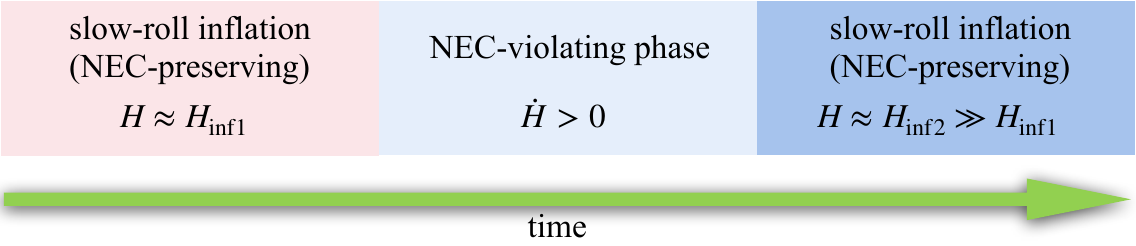}
\caption{In our scenario, the universe begins with a period of slow-roll inflation, and then transitions into a second stage of slow-roll inflation with a higher energy scale, after passing through a phase of violating the null energy condition.} \label{fig-BG230114}
\end{figure}

In this scenario, the universe transits from a first stage of slow-roll inflation with a Hubble parameter $H = H_{\text{inf}1}$, to a second stage of slow-roll inflation with $H = H_{\text{inf}2}\gg H_{\text{inf}1}$, through an intermediate NEC violation stage (see Fig. \ref{fig-BG230114} for an illustration). The NEC violation is able to naturally boost the Hubble parameter $H$ and consequently the power spectrum.
We have constructed the background evolution of such a scenario in \cite{Cai:2020qpu} and investigated the resulted enhanced power spectrum of the primordial gravitational waves (GWs) in \cite{Cai:2022nqv,Cai:2022lec}. Since the current bound of primordial GWs at the CMB band indicates a tensor-to-scalar ratio $r_{0.002} \leq 0.035$ at $95\%$ confidence level \cite{BICEP:2021xfz}, the rich phenomenology of our scenario occurs mainly on smaller scales, including the observational windows of Pulsar Timing Array (PTA) and space-borne GW detectors.

In this letter, we demonstrate that the NEC violation can significantly amplify the abundance of PBHs in single-field inflation through an intermediate violation of the NEC, offering valuable insights into the NEC violation during inflation. Moreover, we investigate the signals of scalar-induced gravitational waves (SIGWs) arising from the amplified density perturbations and show that our results are consistent with current observational constraints. Our findings present a compelling case for the study of PBHs, SIGWs and the primordial GWs as crucial probes for understanding the NEC violation in the very early universe.




{\it Our mechanism.---} The ``no-go'' theorems \cite{Libanov:2016kfc,Kobayashi:2016xpl} indicates that an NEC violation will generically lead to ghost or gradient instabilities in cosmology constructed by the Horndeski theory. In view of that, we should realize our scenario with theories beyond Horndeski \cite{Cai:2016thi,Creminelli:2016zwa,Cai:2017tku,Cai:2017dyi,Kolevatov:2017voe,Ilyas:2020qja}. For simplicity, we will work with the effective field theory (EFT) action
\be
\label{action-230112-1}
S=\int d^4x\sqrt{-g}\Big[\frac{M_{\rm P}^2}{2} {\sf R} + P(\phi,X) + L_{\delta g^{00} {\sf R}^{(3)}} \Big] \,,
\ee
where $X=\nabla_\mu\phi\nabla^\mu\phi$, the EFT operator $L_{\delta g^{00} {\sf R}^{(3)}}=\frac{f(\phi)}{2} \delta g^{00} {\sf R}^{(3)}$ is adopted to thoroughly eliminate the instabilities, $\delta g^{00}$ is the perturbation of the $00-$th component of the metric, ${\sf R}^{(3)}$ is the 3-dimensional Ricci scalar on the spacelike hypersurface, see, e.g., \cite{Cai:2017dyi} for details.

\begin{figure}[htbp]
\includegraphics[scale=2,width=0.85\linewidth]{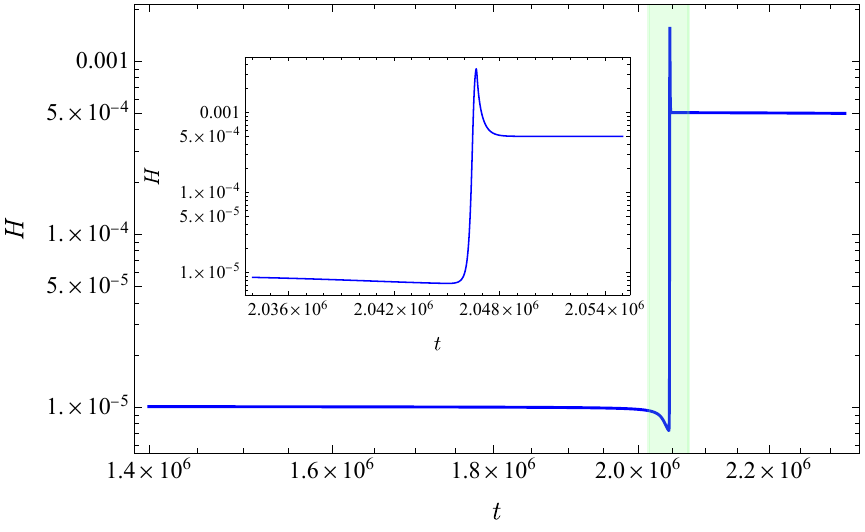}
\caption{A numerical solution of the Hubble parameter $H$ with respect to time $t$ in our model, which results in the blue curve of $P_\zeta$ in Fig. \ref{figPs01}. We have set the Planck scale $M_\text{P}=1$.} \label{figH230515}
\end{figure}

The operator $L_{\delta g^{00} {\sf R}^{(3)}}$ is irrelevant to the background dynamics \cite{Cai:2016thi}. Therefore, the background evolution is determined by the k-essence action
\be
\label{230118P-1}
P(\phi,X) = - {g_1(\phi)\over 2} M_{\rm P}^2 X + {g_2(\phi)\over 4}X^2 -M_{\rm P}^4 V(\phi)\,,
\ee
{where the details are presented in the supplemental material}. The background dynamics of our scenario is illustrated in Fig. \ref{fig-BG230114}. The evolution of the Hubble parameter $H$ is displayed in Fig. \ref{figH230515} by setting a set of parameters. The NEC violating phase can be defined by $\dot{H}>0$.

The quadratic action of scalar perturbation for the action (\ref{action-230112-1}) can be written as
\be
S_{\zeta}^{(2)}=\int d^4x a^3 Q_s\lf[\dot{\zeta}^2-c_s^2
{(\partial\zeta)^2\over a^2} \rt] \,,\label{scalar-action-230112}
\ee
where
\be Q_s={2{\dot \phi}^4P_{XX}-M_{\rm P}^2{\dot H}\over H^2},\quad
c_s^2={M_{\rm P}^2 \over Q_s}\lf({{\dot c}_3\over a} -1\rt)\label{cs2}
\ee
and $c_3=a(1+{2f\over M_{\rm P}^2})/H$, see, e.g., \cite{Cai:2016thi}.
{Obviously, the ghost instability (i.e., $Q_s<0$) and the gradient instability (i.e., $c_s^2<0$) can be easily cured with appropriate construction of $P(\phi,X)$ and the EFT operator $L_{\delta g^{00} {\sf R}^{(3)}}$.}

The equation of motion for $\zeta$ can be written as
\ba
v_k''+\lf({c}_{s}^2
k^2-{z_s''\over z_s} \rt)v_k=0\,, \label{230117eom-us}
\ea
where $v_k=z_s\zeta$ and $z_s=\sqrt{2a^2 Q_s}$, $'\equiv d/d\tau$, $d\tau=a^{-1}dt$. In the following, we will choose specific model parameters in $L_{\delta g^{00} {\sf R}^{(3)}}$ such that the sound speed is canonical, i.e., $c_s^2 \equiv 1$. The perturbation mode is in the vacuum state initially, i.e., $v_k\simeq {1\over \sqrt{2k}}e^{-ik\tau}$. The resulting spectrum of $\zeta$ at the radiation domination stage is $P_{\zeta}={k^3\over 2\pi^2}|\zeta|^2$, which is evaluated after the perturbation modes exited their horizons, i.e., $a H /k \gg 1$.

\begin{figure}[htbp]
\centering %
\includegraphics[scale=2,width=0.45\textwidth]{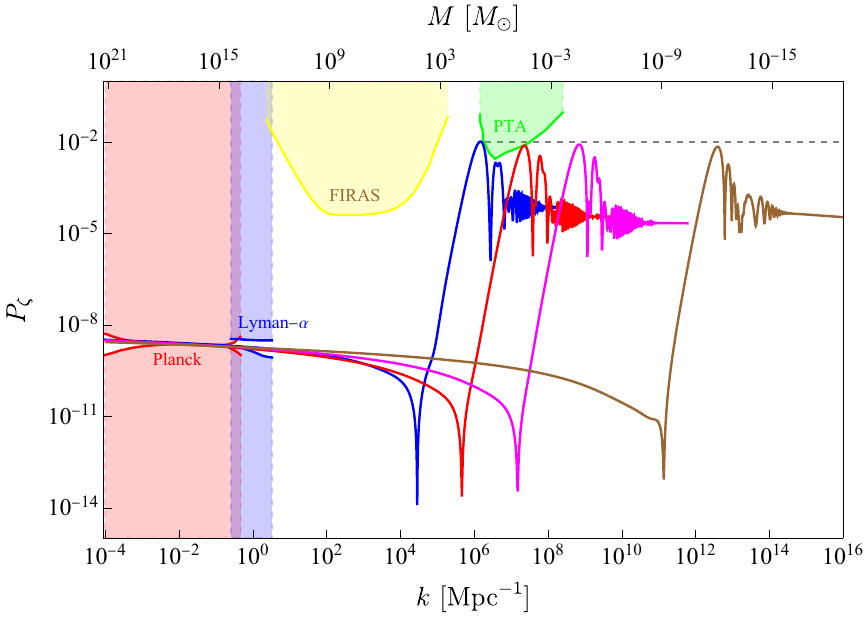}
\caption{The numerical results of the scalar power spectra, $P_\zeta$, are presented for four different parameter sets. The constraints on $P_\zeta$ from Planck, Lyman-$\alpha$, FIRAS, and PTA are depicted as the shadowed regions \cite{Byrnes:2018txb}. The gray dashed line represents the constraint $P_\zeta \simeq 10^{-2}$, which ensures a sufficient abundance of PBHs.}\label{figPs01}
\end{figure}

{Since $c_s^2 \equiv 1$, it can be inferred that the enhancement of the power spectrum is due to the variation of $Q_s$ (primarily the growth of $H$) during the NEC violation, which is intrinsically different from many other mechanisms (including the USR inflation, see the supplemental material for details).} The scalar power spectrum is illustrated in Fig. \ref{figPs01} by numerically solving the background evolution and Eq. (\ref{230117eom-us}) with four different sets of parameters. It should be noted that, for illustrative purposes, the blue and red curves in Fig. \ref{figPs01} are chosen to narrowly satisfy the PTA constraint.

{\it Primordial black holes.---} In this study, we adopt the standard paradigm of PBH formation, where PBHs originate from the gravitational collapse of overdense regions in the early universe. We define the density contrast as $\delta \equiv \delta \rho / \bar{\rho}$, where $\bar{\rho}$ is the energy density at the background level, and $\delta \rho \equiv \rho - \bar{\rho}$ is the density fluctuation. Moreover, we assume that the comoving curvature perturbation $\zeta$ and the density contrast $\delta$ follow a Gaussian distribution. In Fourier space, we have
\begin{equation}
    \delta_k = \frac{2}{3} \left( \frac{k}{aH} \right)^2 \Phi_k \simeq \frac{4}{9} \left( \frac{k}{aH} \right)^2 \zeta_k \,,
\end{equation}
where $\Phi$ is the Bardeen potential in the Newtonian gauge, $\Phi \simeq \frac{2}{3} \zeta$ on super-horizon scales.
Therefore, the power spectrum of density contrast is
\begin{equation}
	\label{eq:Pdelta}
	P_{\delta}(k) = \frac{16}{81} \left( \frac{k}{aH} \right)^4 P_{\zeta}(k) ~.
\end{equation}

In the standard Press-Schechter formalism, the mass fraction function $\beta(M)$, defined as {the fraction of PBHs compared to the total energy of the universe} at the formation time $t_i$, is given by
\begin{equation}
\label{eq:betaPS}
\beta (R) \simeq \frac{\sigma_R}{\sqrt{2\pi} \delta_c} e^{-\frac{\delta_c^2}{2\sigma_R^2}}\,,
\end{equation}
where we assume a Gaussian distribution function for density fluctuations, and $\sigma_R$ represents the corresponding variance. The suggested threshold for PBH formation is $0.4 \leq \delta_c \leq 0.7$ \cite{Musco:2020jjb}. In our case, we will take $\delta_c = 0.5$.

The smoothed density field $\delta_R$ is defined as $\delta_R (\vec{x}) \equiv \int d^3y W(\vec{x} - \vec{y};R) \delta (\vec{y})$,
where $W$ is a window function associated with a characteristic length scale $R \equiv k^{-1}$. We choose the spherically symmetric real-space top-hat window function, i.e., $W(k;R) \equiv {3\left[ \sin (kR) - kR \cos(kR) \right]}{(kR)^{-3}}$,
since it requires the smallest amplitude of density perturbations for a fixed PBH abundance compared to alternative choices \cite{Ando:2018qdb}.

We have $\sigma_R^2 \equiv \langle \delta_R^2\rangle$, where $\langle \delta_R^2\rangle$ is suggested to be \cite{Blais:2002gw,Josan:2009qn}
\begin{equation}
 \langle \delta_R^2\rangle = \int_0^{\infty} \frac{dk}{k} W^2 \frac{16}{81} (kR)^4 T^2 (k, \tau = R)P_{\zeta}(k) ~,
\end{equation}
the scalar transfer function at the radiation dominated era can be given by
\begin{equation}
    T(k,\tau) \equiv \frac{9\sqrt{3}}{(k\tau)^3} \left[ \sin \left( \frac{k\tau}{\sqrt{3}} \right) - \frac{k\tau}{\sqrt{3}} \cos \left( \frac{k\tau}{\sqrt{3}} \right) \right] ~.
\end{equation}

The mass of the PBH is related to the wavenumber $k$ by \cite{Sasaki:2018dmp}
\begin{equation}
	\label{eq:MPBH}
	\frac{M}{M_\odot} \simeq  \left( \frac{\gamma}{0.2} \right) \left( \frac{g_{\ast}}{10.75} \right)^{-\frac{1}{6}} \left( \frac{k}{1.9 \times 10^6 \textnormal{Mpc}^{-1}} \right)^{-2} ~,
\end{equation}
where $M_{\odot}$ is the solar mass, $\gamma$ represents the collapsing efficiency and $g_{\ast}$ denotes the effective number of degrees of freedom for the energy density at PBH formation. In this letter, we take $\gamma = 0.2$ and $g_{\ast}$ = 106.75 \cite{Carr:1975qj}. Accordingly, the current energy fraction, $f_{\textnormal{PBH}}(M)$, is given by \cite{Sasaki:2018dmp}
\begin{equation}
    f_\text{PBH}(M) = \frac{\beta(M)}{2.70 \times 10^{-8}} \left( \frac{k}{1.9 \times 10^6 \textnormal{Mpc}^{-1}} \right)^{-2} ~.
\end{equation}

\begin{figure}[htbp]
\centering %
\includegraphics[scale=2,width=0.45\textwidth]{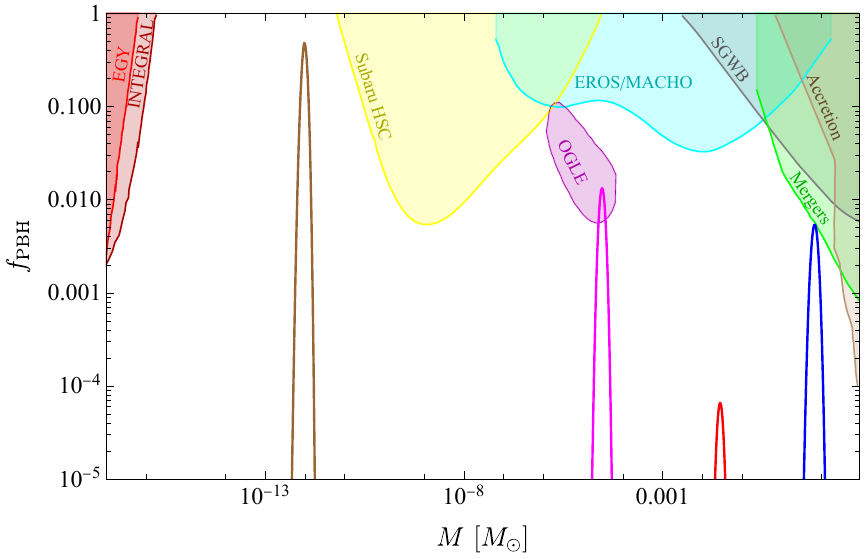}
\caption{The current energy fractions of PBHs, $f_\text{PBH}$, are shown for different parameter sets. The curves of $f_\text{PBH}$ correspond to those of $P_{\zeta}$ with the same color as $P_{\zeta}$ in Fig. \ref{figPs01}. The current constraints on PBH abundance are adopted from \cite{Carr:2020gox}, and the purple shaded region represents the PBH abundance inferred by the OGLE result \cite{Niikura:2019kqi}.}\label{figfPBH01}
\end{figure}

We present the plot of $f_\text{PBH}$ as a function of PBH mass in Fig. \ref{figfPBH01}. The production of PBHs is efficient across various mass scales, as indicated by the brown, magenta, and blue curves. Notably, our model can successfully account for the OGLE ultrashort-timescale microlensing events with specific parameter choices.
Additionally, within the mass range between the red and blue curves (approximately $4\times10^{-2} M_{\odot} \sim 0.8 M_{\odot}$) in Fig. \ref{figfPBH01}, the PBH abundance predicted by our model is constrained to be $f_\text{PBH}<1\%$ due to the PTA constraint on $P_\zeta$, as illustrated by the red and blue curves in Fig. \ref{figPs01}.

Recently, it was claimed that the PBH formation from single-field inflation is ruled out because the enhanced density perturbations give too large one-loop correction to those on CMB scales in the USR inflationary scenario \cite{Kristiano:2022maq}, while its validity is still in debate (e.g., \cite{Fumagalli:2023hpa} pointed out a problem in the computation of \cite{Kristiano:2022maq} and that the large one-loop correction disappears, see also \cite{Riotto:2023gpm}).
Our scenario might be promising to give a loophole for the argument of this problem, due to the intrinsic differences between the NEC violation and the USR mechanism. In our scenario, the enhancement of $P_\zeta$ results primarily from the growth of $H$ instead of the decrease of $\dot{H}$ (or equivalently a very tiny $\epsilon\ll1$ as in the USR inflation). In fact, we have $|\epsilon|\gg1$ during the NEC-violating phase.
Additionally, in the USR scenario, the coefficient of the dominant term in the cubic action and $\zeta$ almost simultaneously reach their maximum values at the end of USR. In contrast, in our scenario, when the coefficient functions in the cubic action reach their maximum values, $\zeta$ or its derivatives are far from reaching their maxima.
Consequently, the argument of \cite{Kristiano:2022maq} does not directly apply to our scenario. Using the approach outlined in \cite{Kristiano:2022maq}, we computed the one-loop corrections to the CMB scale power spectrum for the spectra shown in Fig. \ref{figPs01}, as detailed in the Supplemental Material. The results indicate that our scenario may offer a potentially novel approach to yielding smaller one-loop corrections to the CMB scale power spectrum \footnote{See Supplemental Material, which includes Refs. \cite{Zhu:2021whu,Gao:2011qe,DeFelice:2011uc,Gao:2012ib,Akama:2019qeh}, for additional information about the details of our model, the enhancement of curvature perturbations from the NEC violation, and a discussion of one-loop correction in our scenario.}.
However, since we have made some simplifications in the calculations, obtaining conclusive proof that the result in \cite{Kristiano:2022maq} does not invalidate our scenario necessitates further investigation in the future.



{\it Scalar-induced Gravitational Waves.---} In order to generate a significant abundance of PBHs, it is necessary to enhance the primordial curvature perturbation. This enhancement can potentially lead to the production of large SIGW signals (see e.g., \cite{Domenech:2021ztg} for a review). Hence, it is crucial to examine the corresponding SIGWs within our model to ensure self-consistency. In our analysis, we adopt the standard approach, where the SIGWs are generated as the scalar perturbation modes re-enter the horizon during the radiation domination era.

The power spectrum for SIGW is \cite{Kohri:2018awv}
\begin{align}
    & \ \ \ \ P_h(\tau,k) \nonumber = 576 \int_0^{\infty} dt \int_{-1}^1 ds P_{\zeta} \left(k\frac{t+s+1}{2} \right) \\
    & \nonumber \times P_{\zeta}\left(k\frac{t-s+1}{2} \right) \frac{[-5+s^2+t(2+t)]^4}{(1-s+t)^6(1+s+t)^6} \\
    & \nonumber \times \Bigg\{ \left[ \frac{s^2 - (t+1)^2}{-5+s^2+t(2+t)} + \frac{1}{2} \ln \Big| \frac{-2+t(2+t)}{3-s^2} \Big| \right]^2 \\
    & + \frac{\pi^2}{4}  \Theta (t-\sqrt{3} +1) \Bigg\} ~,
\end{align}
where $\Theta$ is the Heaviside function. It is related to the energy density parameter per logarithmic interval of $k$, $\Omega_\text{GW}(\tau,k)$, as
\begin{equation}
    \Omega_\text{GW}(\tau_r,k) = \frac{\bar{P_h} (\tau,k)}{24} \left( \frac{k}{a(\tau_r)H(\tau_r)} \right)^2  = \frac{\bar{P_h} (\tau,k)}{24} ~,
\end{equation}
where we evaluate the energy density at the horizon re-entry ($k = aH$) with a conformal time $\tau_r$. The energy density spectrum today for SIGW is
\begin{equation}
    \Omega_\text{GW}(k) h^2 = 0.83 \left( \frac{g_{\ast}}{10.75} \right)^{-\frac{1}{3}} \Omega_{r,0} h^2 \Omega_\text{GW}(\tau_r,k) ~,
\end{equation}
where $\Omega_{r,0} h^2 \simeq 4.2 \times 10^{-5}$ is the current density parameter of radiation, see also \cite{Domenech:2019quo,Domenech:2020kqm}.

\begin{figure}[htbp]
\centering %
\includegraphics[scale=2,width=0.48\textwidth]{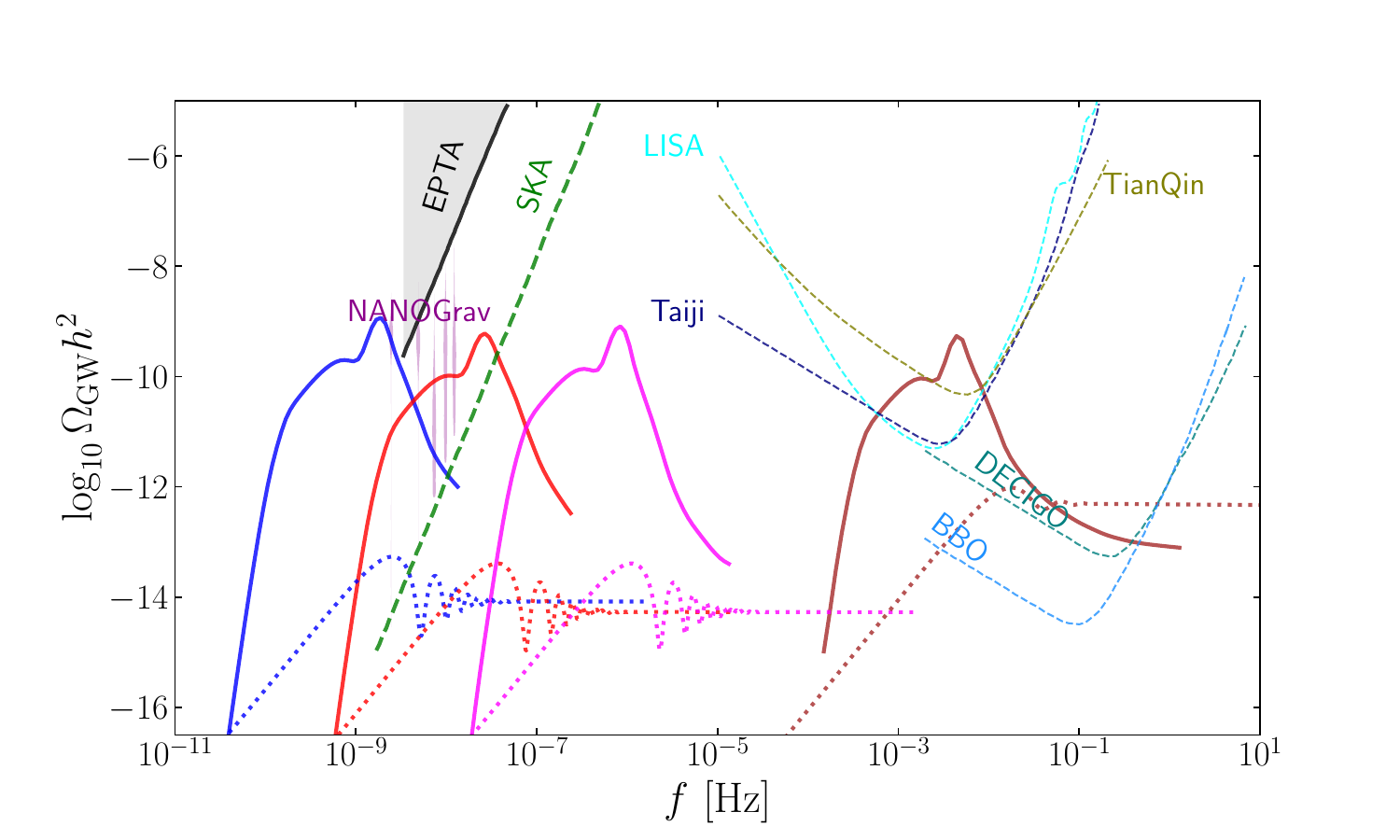}
\caption{The predicted current energy spectra of SIGWs (solid curves) and primordial GWs (dotted curves) are shown for different parameter sets. The curves of $\Omega_\text{GW} h^2$ correspond to those of $P_{\zeta}$ with the same color as shown in Fig. \ref{figPs01}. The shaded region represents the current constraint from EPTA. We also include the expected sensitivity curves of future GW observations as dashed curves, including SKA, LISA, Taiji, TianQin, DECIGO, and BBO. The magenta vertical violin-like bars correspond to the data of NANOGrav.}\label{figIGW01}
\end{figure}

In Fig. \ref{figIGW01}, we present the energy density spectra of SIGWs for the same parameter sets as in Fig. \ref{figPs01} and \ref{figfPBH01}. The SIGW signals corresponding to our parameter sets are consistent with the constraints from the EPTA \cite{Lentati:2015qwp}. Notably, the {solid} blue, red, and brown curves exhibit detectable signatures that fall within the observational windows of future GW detectors, including those of PTA and space-borne GW detectors, see also \cite{Balaji:2022dbi}. Additionally, the red and blue curves may account for the evidence of a stochastic common-spectrum process reported by the NANOGrav Collaboration \cite{NANOGrav:2020bcs}.

{\it Distinctive features of our scenario.---} Basically, the power spectrum of primordial GWs depends primarily on the Hubble parameter $H$, as long as the propagating speed of primordial GWs is $c_T\equiv 1$. In our scenario, $H$ experiences significant growth due to the violation of the NEC, which is unique compared to other single-field PBH formation scenarios, e.g., the USR inflation. Consequently, the resulting primordial GWs spectrum will be significantly enhanced on certain scale and is nearly scale-invariant on smaller scale \cite{Cai:2020qpu}. These distinctive features of primordial GWs are promising to be detected by future observations (e.g., BBO and DECIGO), allowing for distinguishing our NEC violation scenario from the other single-field scenarios of the PBH formation from an observational point of view.

For the four sets of parameter configurations used in Fig. \ref{figPs01}, we have plotted the corresponding primordial GW signals in Fig. \ref{figIGW01} as dotted curves. The blue, red and magenta dotted curves predict $\Omega_\text{GW}h^2\sim 10^{-14}$ on small scales, which is narrowly beyond the sensitivity of the BBO. In contrast, the brown dotted curve predicts $\Omega_\text{GW}h^2\sim 10^{-12}$ within the observation windows of BBO and DECIGO. At this scale and smaller scales, it significantly exceeds the corresponding SIGW signal.
Therefore, synchronized observations of PBHs, SIGWs, and primordial GWs may potentially probe NEC violation during inflation.
Although some of the primordial GW background is borderline detectable in the example cases given, the principle is intriguing and the combination of signals is unique and characteristic for this class of models.


{\it Conclusion and Outlook.---} We report a novel mechanism capable of generating sizable PBHs in the context of single-field inflationary cosmology, by introducing an intermediate  stage of NEC violation during inflation, which offers a unique avenue to enhance the Hubble
parameter $H$ and consequently the primordial power spectrum.
Our scenario can be realized in the EFT framework of inflation.
The primordial curvature perturbation is significantly enhanced in a narrow band of comoving wavelengths corresponding to the NEC violation stage. Consequently, the primordial density perturbation is nearly scale-invariant in both large and small scales connected by a sharp peak.
The sharp peak leads to the generation of a sizable amount of PBHs as well as signals of SIGW, which can be probed and tested in future cosmological surveys.
Furthermore, the distinctive features in the power spectrum of primordial GWs (see \cite{Cai:2020qpu,Cai:2022nqv,Cai:2022lec}) will enable our scenario to be distinguished from other single-field PBH formation scenarios, including USR inflation, from an observational perspective.

In this letter, we adopted an EFT approach in the context of single-field inflationary cosmology, where we assumed a sound speed $c_s\equiv1$ for the scalar perturbations by introducing the EFT operator $L_{\delta g^{00} {\sf R}^{(3)}}$. As a result, our scenario effectively avoids issues related to large entropy fluctuations and super-luminality.
This choice allows for independent parameterization of scalar and tensor perturbations, as the EFT operator does not contribute to the background dynamics or tensor perturbations at quadratic order.
However, in realistic cosmological scenarios derived from covariant actions, scalar and tensor perturbations are interconnected.
For instance, when implementing the EFT operator using theories beyond Horndeski, the sound speeds of scalar and tensor perturbations can be modified. Therefore, to comprehensively investigate the contributions of our scenario to the GW background and confront them with observations, it is necessary to specify the covariant actions and conduct a detailed analysis in future studies.


%


Our work highlighted the potential of utilizing a combination of PBHs, SIGW signals, and primordial GWs as a valuable probe for exploring the NEC violation during inflation, particularly in the era of multi-messenger and multi-band observations.
\\


We thank Chao Chen, Jie-Wen Chen, Chunshan Lin, Taotao Qiu and Shengfeng Yan for valuable discussions.
Y. C. is supported in part by the National Natural Science Foundation of China (Grant No. 11905224), the China Postdoctoral Science Foundation (Grant No. 2021M692942) and Zhengzhou University (Grant No. 32340282).
M. Z. is supported by grant No. UMO 2021/42/E/ST9/00260 from the National Science Centre, Poland.
Y.-S. P. is supported by the National Natural Science Foundation of China under Grant No. 12075246
and the Fundamental Research Funds for the Central Universities.
We acknowledge the use of the computing server {\it Arena317}@ZZU.

\bibliographystyle{utphys}

\bibliography{Ref}

\clearpage

\widetext
\begin{center}
\textbf{\large Supplemental Material for ``Primordial black holes from null energy condition violation during inflation"}
\end{center}
\setcounter{equation}{0}
\setcounter{figure}{0}
\setcounter{table}{0}
\setcounter{page}{1}
\makeatletter
\renewcommand{\theequation}{S\arabic{equation}}
\renewcommand{\thefigure}{S\arabic{figure}}


\section{1. Details of our model}
\label{app:model}

{
Following Ref. \cite{Cai:2020qpu} of the manuscript, we set
\begin{equation}
	g_1(\phi)  =
	-{f_1 e^{2\phi} \over 1+f_1
		e^{2\phi} }+{2\over 1+e^{-q(\phi-\phi_0)}}+{1\over 1+e^{q(\phi-\phi_3)}} ~,
\end{equation}
\begin{equation}
	g_2(\phi) = {f_2\over 1+e^{-q(\phi-\phi_3)}} {1\over
		1+e^{q(\phi-\phi_0)}} ~,
\end{equation}
\be
	V(\phi)  = \Lambda^4 \tanh ^{2}\left({\phi \over \sqrt{6 \alpha } }\right) {1\over
		1+e^{q(\phi-\phi_2)}}
	 +  \lambda \left[1-\frac{(\phi-\phi_1)^{2}}{\sigma^{2}}\right]^{2}{1\over
		1+e^{-p(\phi-\phi_1)}} ~,
\ee
where the model parameters $\lambda$, $\Lambda$, $\alpha$, $\sigma$, $f_{1,2}$, $p$ and $q$ are assumed to be positive constants.
We also assume that $\phi_3<\phi_2<0<\phi_1<\phi_0$, where $\phi_{0,1,2,3}$ are constants.

In the regime of $\phi\ll \phi_3$, we have $g_1= 1$, $g_2= 0$ and
$V=V_\text{inf1}\simeq  \Lambda^4 \tanh ^{2}({\phi / \sqrt{6 \alpha } })$, which can be responsible for setting the first stage of slow-roll inflation and generating primordial curvature perturbations that is consistent with the CMB observations.
For $\phi\gg \phi_0$, we
have $g_1= 1$, $g_2= 0$ and $V=V_\text{inf2}\simeq \lambda
[1-\frac{(\phi-\phi_1)^{2}}{\sigma^{2}}]^{2}\gg V_\text{inf1}$, which results in the second stage of slow-roll inflation.
The scalar field $\phi$ climbs the potential $V(\phi)$ in the intermediate phase with the help of $\phi$-dependent functions $g_1$ and $g_2$ (i.e., noncanonical kinetic term). The parameter values used in the numerical calculations are displayed in Tabel. \ref{tab0817}.

\begin{table}[!bth]
	\begin{tabular}{l|ccccccc}
		\hline\hline
		~\qquad & $~10^7\lambda~$ & $~10^3\Lambda~$ & $\quad\alpha\quad$ & $\quad\sigma\quad$ & $\quad p \quad$& $10^{-3}f_2$\\
		\hline
		Brown & $500$ & $3.4$ & $0.94$ & $11.2$ & $3$ & $2$ \\
		Magenta & $5$ & $3.7$ & $0.9$ & $34$ & $2$ & $4$ \\
		Red & $5$ & $3.92$ & $0.9$ & $38$ & $2$ & $4$  \\
		Blue & $7.7$ & $4.22$ & $0.9$ & $38.9$ & $2$ & $4$  \\
		\hline\hline
	\end{tabular}
	\caption{The parameter values used to produce the brown, magenta, red and blue curves in Fig. 2 to 5. We have set $f_1=1$, $q=4$, $\phi_0=3.2$, $\phi_1=1.6$, $\phi_2=-4$ and $\phi_3=-4.38$ for all curves.}
	\label{tab0817}
\end{table}

\section{2. Enhancement of curvature perturbations}
\label{app:enhance}

Basically, in the single-field slow-roll inflation scenario, the power spectrum of primordial curvature perturbations (i.e., $P_\zeta$) relies on the Hubble parameter $H$, the parameter $\epsilon\equiv-\dot{H}/H^2$ (or its generalized formulation), and the sound speed $c_s$. In the ultra-slow-roll (USR) scenario, $P_\zeta$ is enhanced primarily by a very small value of $\epsilon\ll1$. In contrast, in our NEC violation scenario, $P_\zeta$ is enhanced primarily by the growth of $H$. This point can be demonstrated as follows.

As shown by the blue curve in the zoomed-in figure of Fig. 2, the maximum value (i.e., the peak value) of the Hubble parameter is $H_\text{max}\approx3.5\times10^{-3}$, while the initial value of $H$ is $H_\text{ini}\approx1\times10^{-5}$. Hence, we have
\be H_\text{max}^2/H_\text{ini}^2\approx1.2\times10^5\,.\ee
The maximum value (i.e., the peak value) of $P_\zeta$ is $P_{\zeta,\text{max}}\approx1\times10^{-2}$, whereas  $P_\zeta=P_{\zeta,\text{inf1}}\approx2\times10^{-9}$ for the modes generated by the first inflationary stage. Therefore, $P_\zeta$ experiences an enhancement by a factor of
\be P_{\zeta,\text{max}}/P_{\zeta,\text{inf1}}\approx5\times10^6\,.
\ee
{We can see that $P_\zeta$ undergoes a growth of about $6$ to $7$ orders of magnitude, with approximately $5$ orders of magnitude attributed to the contribution from the growth of $H^2$ and the remaining $1$ order of magnitude attributed to contribution from the transition phase. Therefore, the growth of $P_\zeta$ should be primarily attributed to the growth of $H^2$. The contribution from the transition phase is secondary.}

A similar analysis can be applied to the brown curve in Fig. 3. The maximum value of the Hubble parameter is approximately $H_\text{max}\approx5.8\times10^{-3}$, while the initial value of $H$ is $H_\text{ini}\approx6.6\times10^{-6}$. This yields
\be
H_\text{max}^2/H_\text{ini}^2\approx7.8\times10^5\,.
\ee
The maximum value of $P_\zeta$ is $P_{\zeta,\text{max}}\approx7\times10^{-3}$, whereas $P_\zeta=P_{\zeta,\text{inf1}}\approx2\times10^{-9}$ for the modes generated during the first inflationary stage. Consequently, $P_\zeta$ undergoes an enhancement by a factor of
\be
P_{\zeta,\text{max}}/P_{\zeta,\text{inf1}}\approx3.5\times10^6\,,
\ee
which should be primarily attributed to the growth of $H^2$ as well.
{Compared to the blue curve, for the brown curve, the contribution of the transition phase to the growth of $P_\zeta$ is slightly smaller.}
Similar conclusions can also be drawn for the red and magenta curves.

The dynamics of primordial curvature perturbations are complex in a generic NEC violation phase.
Analytically, the evolution of the curvature perturbations in an NEC violation phase is studied in the context of bouncing cosmology (see, e.g., \cite{Zhu:2021whu}), in which the bouncing phase also violates the NEC. For the purpose of illustration, we consider a simplified situation where the NEC violation phase is short enough such that the Hubble parameter $H$ can be approximately treated as a linear function of cosmic time $t$, i.e., $H = \Gamma t$,  with $\Gamma$ being a constant. In this case, the time derivative of scalar field can be parameterized as $\dot{\phi} \propto e^{-t^2}$, and the corresponding mode function of scalar perturbation on super-horizon mode is
\begin{equation}
	v_k(\tau) \simeq b_1(k) e^{\sqrt{\Gamma} \tau} + b_2(k)e^{-\sqrt{\Gamma} \tau} ~.
\end{equation}
In conclusion, during the NEC-violating phase, the mode function of the super-horizon mode is composed by an exponentially growing part and an exponentially decreasing part.

The dynamics of super-horizon modes depend on the coefficients $b_1$ and $b_2$, which are determined by the matching conditions. When $b_1 \gg b_2$, the curvature perturbation experiences exponential enhancement. Hence, in this simplified scenario, we can deduce that the super-horizon curvature perturbation modes during the NEC-violating phase undergo enhancement due to the dominance of the exponentially growing component.

{Regarding the rapid growth of $H$ as the cause of perturbation growth outside the horizon, we can also understand it as follows.
In the super-horizon limit (i.e., $k^2\ll z_s^{\prime\prime}/z_s$), the perturbation equation simplifies to $v_k^{\prime\prime}-v_k z_s^{\prime\prime}/z_s=0$. It is well known that in this regime, the solution for the primordial curvature perturbation mode can be decomposed into a constant mode and an evolving mode, i.e.,
\begin{equation}
    \zeta \equiv v_k/z_s = \zeta_c + \zeta_e ~,\quad~ \zeta_c = {\rm const} ~,\quad~ \zeta_e \propto \int \frac{d\tau}{z_s^2} ~ \to ~ |\dot{\zeta}| = \frac{1}{a} |\zeta^{\prime}| = \frac{D_k}{a z_s^2} ~,\label{eq:zetace001}
\end{equation}
where $z_s=\sqrt{2a^2 Q_s}$, $D_k$ is a $k$-dependent integration constant. A comparison between the pure numerical solution of $|\zeta'|$ and the analytical approximation provided by Eq. (\ref{eq:zetace001}) is shown in Fig. \ref{fig:Comp-zetap002} for the parameter set corresponding to the brown curve in Fig. 3 of the main text.

During slow-roll inflation, we have $H\approx \text{const.}$ and $z_s\simeq a\sqrt{\epsilon}\simeq \sqrt{\epsilon}/(-H\tau)$, which indicates that $\zeta_e \propto H^2|\tau|^3/\epsilon$. Therefore, after perturbations exit the horizon during inflation, the evolving mode $|\zeta_e|$ rapidly decreases and becomes much smaller than the constant mode $\zeta_c$.

During NEC violation, $\dot{H}>0$, we roughly have $|\epsilon|\gg1$, and $H$ increases rapidly with time, while the scale factor $a$ grows insignificantly. During a certain period within this timeframe, the non-canonical kinetic terms in the background equations dominate over the potential term. Consequently, we have $g_1(\phi)X\approx 12H^2+6\dot{H}$ and $g_2(\phi)X^2\approx 12H^2+4\dot{H}$. We find $Q_s\simeq(12-3\epsilon)M_\text{P}^2\approx 3|\epsilon|M_\text{P}^2$.
The rapid growth of $H$ (whether exponential or power-law) will lead to a rapid decrease of $Q_s\simeq 3|\epsilon|$. Therefore, $\zeta_e'\propto z_s^{-2}\propto Q_s^{-1}$ will also increase dramatically as $H$ grows rapidly. After the end of NEC violation, the change in $Q_s$ is no longer significant, while $a$ grows exponentially, causing $\zeta'\propto a^{-2}Q_s^{-1}$ to decrease. As a result, the evolution of $\zeta'$ around NEC violation resembles a $\delta$-function, as shown in Fig. \ref{fig:Comp-zetap002} (a). After integrating with respect to $\tau$, this will evidently lead to a rapid growth of $|\zeta_e|$.

If we define the integrals of the rising and falling parts of $\zeta'$ in the left and right halves of Fig. S1 (a) as $\Delta \zeta_1=\int_{\tau_\text{left}}^{\tau_\text{mid}} \zeta' d\tau$ and $\Delta \zeta_2=\int_{\tau_\text{mid}}^{\tau_\text{right}} \zeta' d\tau$, respectively, then approximately $\Delta \zeta_1$ and $\Delta \zeta_2$ represent the contributions to the enhancement of $P_\zeta$ from the rapid growth of $H$ and the transition, respectively. Here, $\tau_\text{mid}$ is the time at which $\zeta'$ reaches its maximum value. Since $a(\tau_\text{left}<\tau<\tau_\text{mid}) \ll a(\tau_\text{mid}<\tau<\tau_\text{right})$ and $d\tau=dt/a$, for the $\zeta'$ shown in Fig. S1 (a) (corresponding to the wavenumber $k$ associated with the maximum of the brown power spectrum in Fig. 3), we will find that $\Delta \zeta_1\gg\Delta \zeta_2$. Similar conclusions can be drawn for power spectra associated with other colors. Therefore, the growth of $\zeta$ and thereby the enhancement of $P_\zeta$ should primarily be attributed to the rapid growth of $H$.

NEC violation ($|\epsilon|\gg1$) and USR ($|\epsilon|\ll 1$) both achieve the enhancement of power spectrum at specific scales by allowing the evolving mode $|\zeta_e|$ to grow outside the horizon, leveraging the decrease in $|\epsilon|\equiv|\dot{H}/H^2|$.
The difference lies in how they achieve the rapid decrease in $|\epsilon|$: NEC violation occurs through the rapid increase of $H$, while USR occurs through the rapid decrease of $\dot{H}$.


\begin{figure}[htbp]
\subfigure[]{\includegraphics[width=.48\textwidth]{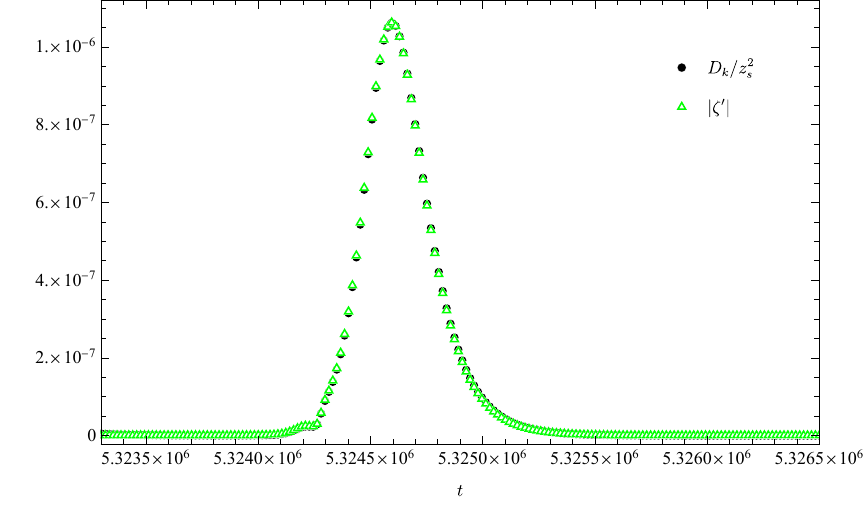} }
\subfigure[]{\includegraphics[width=.47\textwidth]{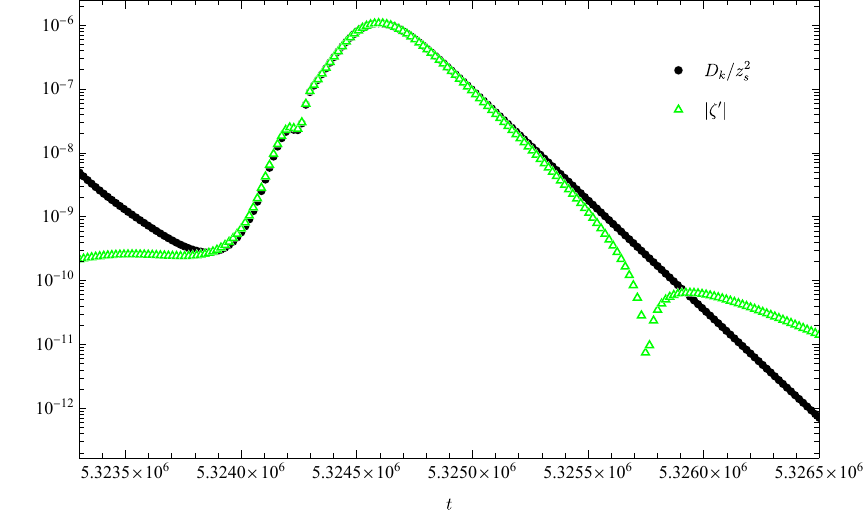} }
\caption{A comparison between the pure numerical solution of $|\zeta'|$ and the analytical approximation provided by Eq. (\ref{eq:zetace001}) is shown for the parameter set corresponding to the brown curve in Fig. 3 of the main text. Here, we consider $k=4\times 10^{12}$ $\text{Mpc}^{-1}$ and set $D_k=1.37\times 10^{24}$. The time range on the horizontal axis is around the NEC violation. In the left panel, we use a conventional scale, while in the right panel, we employ a logarithmic scale. Similar computations can be conducted for other curves presented in Fig. 3 of the main text and for different values of $k$.
 } \label{fig:Comp-zetap002}
\end{figure}

Using the parameter set corresponding to the brown curve in Fig. 3 as an example, we illustrate the evolution of the curvature perturbation $\zeta_k$ numerically by plotting the evolutions of $k^{3/2}|\zeta_k|$ in Fig. \ref{fig:zeta230907}.
\begin{itemize}
\item For modes that cross the horizon during the first inflationary phase, we illustrate the dynamics of $|\zeta_k|$ at a representative scale of $k = 10^{-2} \textrm{Mpc}^{-1}$ in Fig. \ref{fig:zeta230907}(a). We can see that $|\zeta_k|$ decreases within the sub-horizon regime and remains constant after crossing the horizon. Consequently, the power spectrum of these modes is $P_{\zeta} \propto H^2_{\textrm{inf1}}$.

\item For modes cross the horizon during the NEC violation phase, we illustrate the dynamics of $|\zeta_k|$ at a representative scale of $k = 10^{12} \textrm{Mpc}^{-1}$ in Fig. \ref{fig:zeta230907}(b). We can see that $|\zeta_k|$ experiences an exponential amplification on super-horizon scale due to the growth of $H$. After the NEC-violating phase, the curvature perturbation remains invariant on super-horizon scale in the second inflationary phase. The power spectrum is enhanced on corresponding scale accordingly.

\item For modes cross the horizon during the second inflationary phase, we plot $|\zeta_k|$ at a representative scale of $k = 10^{15} \textrm{Mpc}^{-1}$ in Fig. \ref{fig:zeta230907}(c). Obviously, $|\zeta_k|$ decreases during the first inflationary stage and experiences nontrivial evolution due to the nontrivial behavior of $z_s$ during the NEC-violating phase. After exiting horizon in the second inflationary phase, $|\zeta_k|$ becomes frozen and results in a spectrum $P_{\zeta} \propto H^2_{\textrm{inf2}}$.
\end{itemize}

\begin{figure}[htbp]
    \subfigure[]{\includegraphics[width=.45\textwidth]{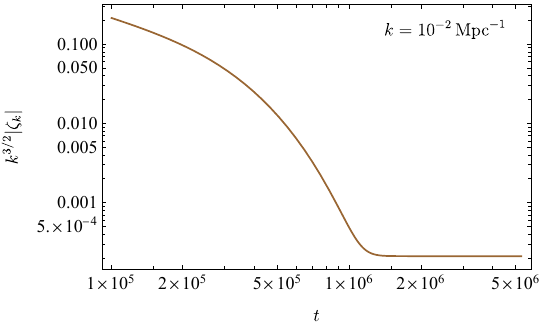} }
    \subfigure[]{\includegraphics[width=.45\textwidth]{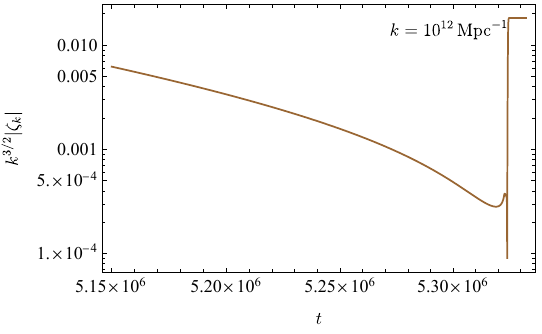} }
    \subfigure[]{\includegraphics[width=.45\textwidth]{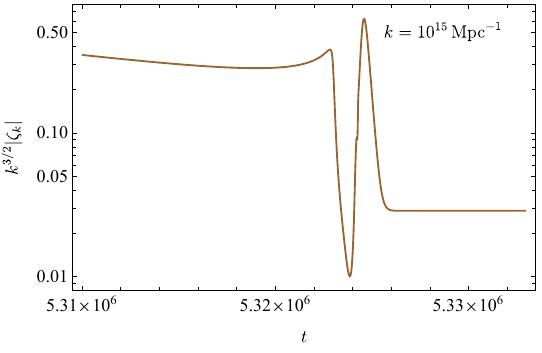} }
    \caption{The evolution of the re-scaled curvature perturbations $|k^{{3/2}}\zeta_k|$ with $k=10^{-2}\,\text{Mpc}^{-1}$, $10^{12}\,\text{Mpc}^{-1}$, and $10^{15}\,\text{Mpc}^{-1}$, which exit the horizon during the first inflationary phase, the NEC-violating phase, and the second inflationary phase, respectively.} 	\label{fig:zeta230907}
\end{figure}

The evolution of the perturbation mode shown in Fig. \ref{fig:zeta230907} can be qualitatively understood from Eq. (\ref{eq:zetace001}) as follows:
\begin{itemize}
	\item For the perturbation modes that exited the horizon during the first stage of slow-roll inflation, their evolving mode $|\zeta_e|$ undergoes significant decay during inflation. Therefore, even though $|\zeta_e|$ starts to increase during the NEC-violating phase, it remains much smaller than $|\zeta_c|$ throughout. Namely, for these large-scale perturbation modes (e.g., $k=10^{-2}$ $\text{Mpc}^{-1}$), $\zeta$ is dominated by the constant mode $\zeta_c$ after exiting the horizon.
	\item For perturbation modes that exit the horizon around the NEC-violating phase (e.g., $k\simeq 10^{12}$ $\text{Mpc}^{-1}$), their evolving mode undergoes growth and may eventually dominate over the constant mode. For this reason, we obtain an enhanced power spectrum at the corresponding scale.
    \item For perturbations that exit the horizon during the second stage of slow-roll inflation, Eq. (\ref{eq:zetace001}) is not applicable around the NEC-violating phase.
\end{itemize}
}

}

\section{3. One-loop correction in our scenario}

Recently, it has been argued that an enhanced small-scale spectrum eligible for an abundant primordial black hole (PBH) formation in the USR scenario will inevitably lead to a significant change to the CMB-scale fluctuation \cite{Kristiano:2022maq}. It is thus interesting to examine whether our scenario for PBH formation is safe under the one-loop correction.

The leading one-loop correction to perturbations on CMB scale is
\begin{equation}
    \langle \zeta(\vec p) \zeta(-\vec p) \rangle = \langle \zeta(\vec p) \zeta(-\vec p) \rangle_{(1,1)} + \left[ \langle \zeta(\vec p) \zeta(-\vec p) \rangle_{(0,2)} + \langle \zeta(\vec p) \zeta(-\vec p) \rangle_{(0,2)}^{\dagger} \right] ~,
\end{equation}
where in the rest of the section, we will use the label $p$ to denote the scale at the CMB range. The ratio of the one-loop correction to the CMB power spectrum is
\begin{equation}
\label{eq:oneloopexpression}
    \Delta P_{\zeta}(p) \equiv \frac{P_{\zeta,1}(p)}{P_{\zeta,0}(p)}  = \frac{\langle \zeta(\vec p) \zeta(-\vec p) \rangle_{(1,1)}}{|\zeta_p(0)|^2} + \frac{2 {\rm Re} \langle \zeta(\vec p) \zeta(-\vec p) \rangle_{(0,2)}}{|\zeta_p(0)|^2} ~,
\end{equation}
which can be evaluated using the in-in formalism
\begin{equation}
    \langle \zeta(\vec p) \zeta(-\vec p) \rangle_{(1,1)} (t) \equiv \int_{-\infty}^{t} dt_1  \int_{-\infty}^{t} dt_2 \langle H_{\rm int} (t_1) \hat{\zeta}(\vec{p})  \hat{\zeta}(-\vec{p}) H_{\rm int} (t_2) \rangle ~,
\end{equation}
\begin{equation}
    \langle \zeta(\vec p) \zeta(-\vec p) \rangle_{(0,2)} (t) \equiv -\int_{-\infty}^{t} dt_1  \int_{-\infty}^{t} dt_2 \langle  \hat{\zeta}(\vec{p})  \hat{\zeta}(-\vec{p}) H_{\rm int} (t_1) H_{\rm int} (t_2) \rangle ~.
\end{equation}

\subsection{3.1. Interacting Hamiltonian}
Our action is composed of a k-essence action and an EFT operator. The perturbed metric, in terms of ADM variables, is
\begin{equation}
\label{eq:ADM}
    N = e^{\alpha} ~,~ N_i = a \partial_i \beta ~,~ h_{ij} = a^2 e^{2\zeta} \delta_{ij} ~.
\end{equation}
In the presence of EFT operator, the expressions for $\alpha$ and $\beta$ are
\begin{equation}
    \alpha = \frac{\zeta^{\prime}}{aH} ~,~ \partial^2 \beta = \frac{Q_s}{M_p^2} \zeta^{\prime} - \frac{\partial^2 \zeta}{aH} \left( 1 + \frac{f(\phi)}{2M_p^2} \right) ~,~ \beta = \frac{Q_s}{M_p^2} \psi - \frac{\zeta}{aH} \left( 1 + \frac{f(\phi)}{2M_p^2} \right)  ~.
\end{equation}
We write the cubic action in the following form
\begin{equation}
    L_{\zeta}^{(3)} = L_{\zeta,P}^{(3)} + \Delta L_{\zeta}^{(3)} ~,
\end{equation}
where $L_{\zeta,P}^{(3)}$ is the cubic Lagrangian of the curvature perturbation for a purely k-essence theory, and the term $\Delta L_{\zeta}^{(3)}$ characterize the modification from the EFT operator. It is easy to see that the modification from EFT operator is proportional to the coupling function, i.e.,
\begin{equation}
    \Delta L_{\zeta}^{(3)} \propto f(\phi) ~,\quad~ L_{\zeta,P}^{(3)} = L_{\zeta}^{(3)} (f = 0) ~.
\end{equation}

The cubic Lagrangian for a k-essence theory is well-known \cite{Gao:2011qe,DeFelice:2011uc,Gao:2012ib, Akama:2019qeh}
\begin{align}
\label{eq:Lkessence}
    L_{\zeta,P}^{(3)} & \nonumber = Q_s \Big\{ \frac{\Lambda_1}{H} \zeta^{\prime 3} + a \Lambda_2 \zeta \zeta^{\prime 2} + a \Lambda_3 \zeta (\partial_i \zeta)^2 + \frac{\Lambda_4}{aH^2} \zeta^{\prime 2} \partial^2 \zeta + a\Lambda_5 \zeta^{\prime} \partial_i \zeta \partial^i \psi + a \Lambda_6 \partial^2 \zeta (\partial_i \psi)^2 \\
    & + \frac{\Lambda_7}{aH^2} [\partial^2 \zeta (\partial_i \zeta)^2 - \zeta \partial^i \partial^j (\partial_i \zeta \partial_j \zeta) ] + \frac{\Lambda_8}{H} [\partial^2 \zeta \partial_i \zeta \partial^i \psi - \zeta \partial^i \partial^j (\partial_i \zeta \partial_j \psi) ] \Big\} + F(\zeta)E_s ~,
\end{align}
where $\psi \equiv \partial^{-2} \zeta^{\prime}$ and the dimensionless coefficients are
\begin{equation}
    \Lambda_1 = \frac{2}{3} - \frac{2\epsilon - \mathcal{R}_s}{6\epsilon} \frac{Q_s^2}{M_p^4} ~,~ \Lambda_2 = \epsilon - \mathcal{R}_s ~,~ \Lambda_3 = 1 + \mathcal{R}_s + \epsilon \left( 1 - \frac{M_p^2}{Q_s} \right) ~,
\end{equation}
\begin{equation}
    \Lambda_4 = \Lambda_7 = \Lambda_8 = 0 ~,~ \Lambda_5 = \frac{Q_s}{M_p^2} \left(  \frac{\epsilon}{2} - 2 \right) ~,~ \Lambda_6 = \frac{Q_s}{4M_p^2}  ~,
\end{equation}
and
\begin{equation}
    \epsilon \equiv - \frac{\dot{H}}{H^2} ~,~ \mathcal{R}_s \equiv \frac{\dot{Q}_s}{HQ_s} = \frac{Q_s^{\prime}}{aHQ_s} ~.
\end{equation}
The last term in Eq. (\ref{eq:Lkessence}), i.e., $F(\zeta)E_s$, represents a pure boundary term, which we will disregard.
Following \cite{Kristiano:2022maq}, we will focus on the dominant term in Eq. (\ref{eq:Lkessence}) when evaluating the one-loop correction.

\begin{figure}[htbp]
    \subfigure[~]{\includegraphics[width=.47\textwidth]{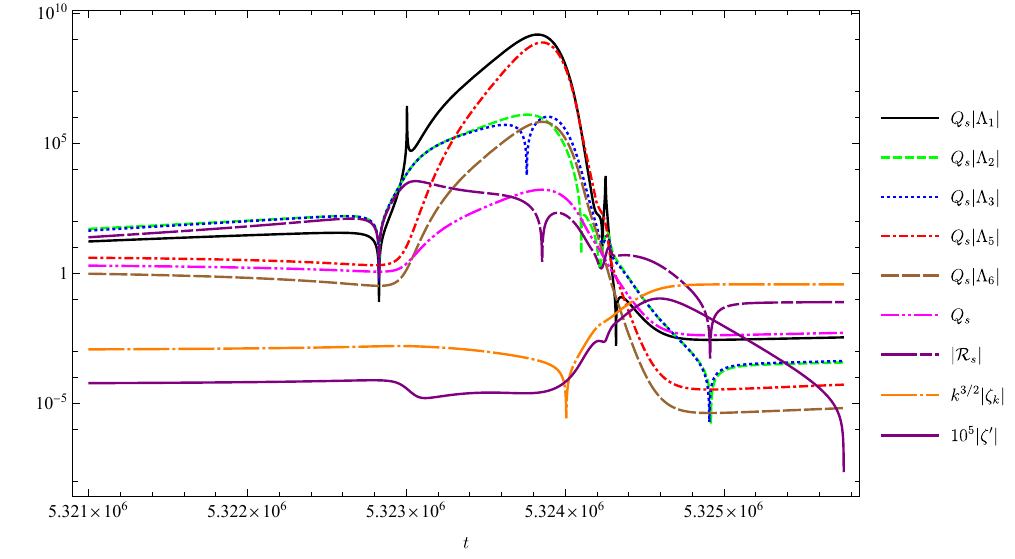} }
    \subfigure[~]{\includegraphics[width=.48\textwidth]{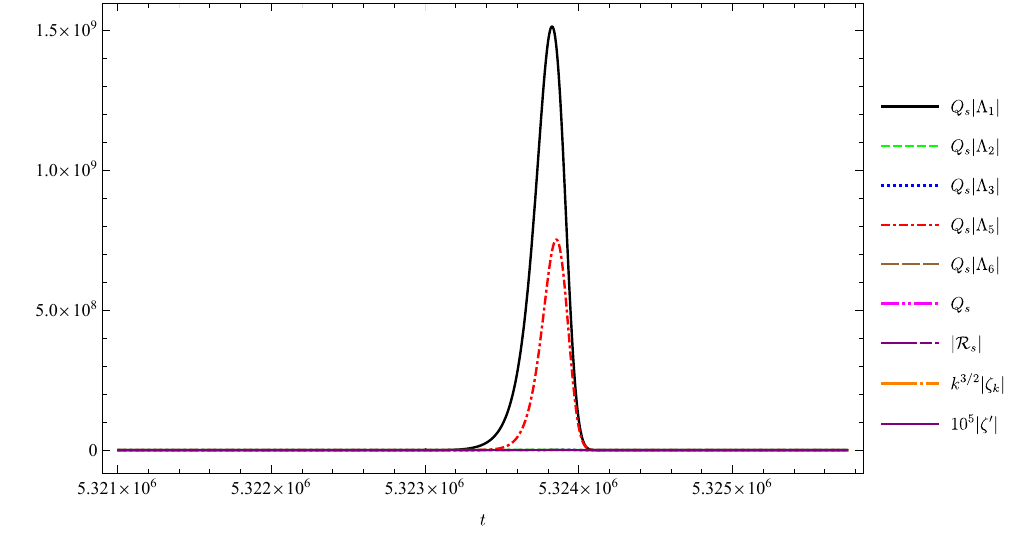} }
    \caption{The evolution of $Q_s|\Lambda_1|$, $Q_s|\Lambda_2|$, $Q_s|\Lambda_3|$, $Q_s|\Lambda_5|$, $Q_s|\Lambda_6|$, $Q_s$, $|{\cal R}_s|$, $k^{3/2}|\zeta_k|$ and $10^5|\zeta'|$. As an example, we set $k=4\times10^{12}$ $\text{Mpc}^{-1}$, which corresponds to the scale where the power spectrum peaks for the brown curve in Fig. 3 of the main text.} \label{fig:QsLambdas}
\end{figure}

We depict in Fig. \ref{fig:QsLambdas} the evolution of the coefficient functions in Eq. (\ref{eq:Lkessence}) around the NEC violation for the parameter set corresponding to the brown curve in Fig. 3 of the main text. These coefficient functions reach their peaks during the NEC violation period and gradually return to the slow-roll approximation during the first and second inflationary stages. We also plot the evolution of the perturbation mode and its derivative in Fig. \ref{fig:QsLambdas}, where we set $k=4\times10^{12}$ $\text{Mpc}^{-1}$, corresponding to the scale where the power spectrum peaks for the brown curve in Fig. 3 of the main text. For clarity, we use logarithmic scale in the left panel and conventional scale in the right panel. Similar calculations can be performed for the magenta, red and blue curves in Fig. 3 of the main text.

It is worth noting that the moments when these coefficient functions reach their peaks are in the middle of the NEC violation stage, during which the perturbation modes (and their derivatives) outside the horizon are far from reaching their peaks. This implies a mismatch between the moments when the coefficients of terms in Eq. (\ref{eq:Lkessence}) reach their maxima and those of $\zeta^3$, $(\zeta^\prime)^3$, etc. This feature is different from the USR scenario, where both the coefficients and perturbation modes reach their maxima near the end of the USR phase.


Following the spirit of \cite{Kristiano:2022maq}, we can infer from Fig. \ref{fig:QsLambdas} that the leading one-loop contribution from \eqref{eq:Lkessence} comes from the $Q_s \Lambda_1$ term. The $Q_s \Lambda_5$ term may also contribute significantly. For simplicity, we primarily compute the one-loop correction of the $Q_s \Lambda_1$ term, assuming that the $Q_s \Lambda_5$ term will contribute roughly the same or even less. This simplification should provide us with useful insights.

Additionally, we need to consider the modification from the EFT operator, denoted as $\Delta L_{\zeta}^{(3)}$, which consists of two cubic terms, i.e.,
\begin{equation}
\label{eq:DeltaL1}
    \Delta L_{\zeta,1}^{(3)} = (f(\phi) R^{(3)} \delta g^{00}/2)^{(3)} = \frac{2 f(\phi) \zeta^{\prime}}{a^3H} [4\zeta \partial^2 \zeta - (\partial \zeta)^2] ~,
\end{equation}
where we have used
\begin{equation}
    \delta g^{00} = 1 + g^{00} \simeq 2\alpha ~,\quad ~ R^{(3)} = -\frac{2}{a^2} e^{-2\zeta} [2\partial^2 \zeta + (\partial \zeta)^2] ~.
\end{equation}
However, the coupling function $\sim f(\phi)$ has a peak value of order $\mathcal{O}(10^3)$. Therefore, the contribution form \eqref{eq:DeltaL1} is subdominant compared to that from \eqref{eq:Lkessence}.

The EFT operator also modifies the shift vector according to \eqref{eq:ADM}, thus the cubic action from the $P(\phi,X)$ term will differ from \eqref{eq:Lkessence}. To account for the modification, we write down the cubic action for a generic k-essence theory without any integration by part:
\begin{eqnarray}
         & & S^{(3)}_{\zeta,P} \nonumber \\
         & = & \int dt d^{3}x~ a^3 \left\{-9\zeta\zeta'^{2}+2\zeta'\left(\zeta\partial^{2}\beta+\partial_{i}\zeta\partial^{i}\beta\right)-\alpha\left(\partial_{i}\zeta\right)^{2}+\left(\partial_{i}\beta\right)^{2}\partial^{2}\zeta-\frac{1}{2}\zeta\left(4\alpha\partial^{2}\zeta-\left(\partial^{2}\beta\right)^{2}+\left(\partial_{i}\partial_{j}\beta\right)^{2}\right)\right.\nonumber \\
         &  & + \zeta\left(\partial_{i}\zeta\right)^{2} - 3\mathcal{H}^{2}(3-Q_s)\alpha^{2}\zeta + 2\mathcal{H} \alpha\left(9\zeta\zeta'-\zeta\partial^{2}\beta-\partial_{i}\zeta\partial^{i}\beta\right)\nonumber \\
         &  & \left.+ \alpha\left[3\zeta'^{2}-2\zeta'\partial^{2}\beta+\frac{1}{2}\left(\left(\partial^{2}\beta\right)^{2}-\left(\partial_{i}\partial_{j}\beta\right)^{2}\right)\right] - \mathcal{H}\alpha^{2}\left(3\zeta'-\partial^{2}\beta\right) - \alpha^{2}\partial^{2}\zeta+\frac{\lambda_{5}}{2}\mathcal{H}^{2}\alpha^{3}\right\} ~.\label{S3_zeta_alpha_beta}
        \end{eqnarray}
The EFT contribution from the modification to the shift vector then reads:
\begin{align}
    \Delta L_{\zeta,2}^{(3)} & \nonumber = L_{\zeta,P}^{(3)} \big(\beta = Q_s \psi/M_p^2 - a^{-1}H^{-1} \zeta \left( 1 + f(\phi)/2M_p^2 \right)\big) - L_{\zeta,P}^{(3)} (\beta = Q_s \psi/M_p^2 - \zeta /aH) \\
    & = \nonumber \frac{f^2}{4aH^2 M_p^4} \left[ \partial^{2}\zeta (\partial_i \zeta)^2 + \frac{1}{2} (\zeta + \alpha) [(\partial^2 \zeta)^2 - (\partial_i \partial_j \zeta)^2]  \right] \\
    & - \frac{Q_s f }{2H M_p^4} \left[ \partial^{2}\zeta \partial_i \zeta \partial_i \psi + (\zeta + \alpha) (\zeta^{\prime} \partial^2 \zeta - \partial_i \partial_j \psi \partial_i \partial_j \zeta)  \right] ~.
\end{align}
The peak values of the coefficients of the modification are on the order of $\mathcal{O}(10^6)$, much smaller than $\mathcal{O}(10^9)$.

Therefore, in the subsequent calculation of the one-loop correction, we will primarily focus on the dominant interacting Hamiltonian around the NEC-violating phase, which is the $Q_s \Lambda_1$ term, i.e.,
\begin{equation}
\label{eq:Hint}
    H_{\rm int} = - \int d^3x a^3 Q_s \Lambda_1 H^{-1} \dot{\zeta}^{3} ~.
\end{equation}
In our setup, the quantity $Q_s$ has dimensions of $[M]^2$, while the curvature perturbation $\zeta$ is dimensionless. The interaction Hamiltonian \eqref{eq:Hint} has the expected mass dimension.

\subsection{3.2. The time integral}

The standard inflation scenario should not predict a large one-loop correction on CMB scale. Therefore, we rewrite the interaction Hamiltonian in the following form:
\begin{equation}
    H_{\rm int} = H_{\rm int,inf} + \Delta H_{\rm int} ~,
\end{equation}
where $H_{\rm int,inf}$ is the interacting Hamiltonian for a canonical inflation scenario, and $\Delta H_{\rm int}$ characterizes the deviation from the canonical scenario. For any operator $\mathcal{P}$, we have for instance
\begin{align}
    \langle \mathcal{P}(t) \rangle_{(1,1)} & \nonumber = \int_{-\infty}^{t} dt_1 \int_{-\infty}^{t} dt_2 \langle H_{\rm int} (t_1) \mathcal{P}(t) H_{\rm int} (t_2) \rangle \\
    & \nonumber = \int_{-\infty}^{t} dt_1 \int_{-\infty}^{t} dt_2 \langle (H_{\rm int,inf} (t_1) + \Delta H_{\rm int} (t_1)) \mathcal{P}(t) (H_{\rm int,inf} (t_2) + \Delta H_{\rm int} (t_2)) \rangle \\
    & = \int_{-\infty}^{t} dt_1 \int_{-\infty}^{t} dt_2 \langle H_{\rm int,inf} (t_1)  \mathcal{P}(t) (H_{\rm int,inf} (t_2) \rangle \label{eq:infcontribution} \\
    & + \int_{-\infty}^{t} dt_1 \int_{-\infty}^{t} dt_2 \langle \Delta H_{\rm int} (t_1) \mathcal{P}(t)  \Delta H_{\rm int} (t_2) \rangle \label{eq:NECVcontribution} \\
    & \label{eq:crosscontribution} + \int_{-\infty}^{t} dt_1 \int_{-\infty}^{t} dt_2 \left[ \langle \Delta H_{\rm int} (t_1) \mathcal{P}(t) H_{\rm int,inf} (t_2) \rangle + \langle H_{\rm int,inf} (t_1) \mathcal{P}(t) \Delta H_{\rm int} (t_2) \rangle \right] ~.
\end{align}
The term \eqref{eq:infcontribution} characterizes the one-loop correction from a canonical slow-roll inflation action, which shall be negligible. The time integral \eqref{eq:NECVcontribution} and \eqref{eq:crosscontribution} are significant only around the NEC violation. Furthermore, the magnitude of \eqref{eq:NECVcontribution} is much larger than that of \eqref{eq:crosscontribution} around the NEC violation. Hence, we will only consider the \eqref{eq:NECVcontribution} term.

In order to simplify the calculation, based on the numerical result shown in Fig. \ref{fig:QsLambdas} (especially in Fig. \ref{fig:QsLambdas} (b)), we parameterize the interacting Hamiltonian as a delta function, i.e.,
\begin{equation}
    Q_s \Lambda_1 = A \delta (t-t_0) ~,\quad~ \Delta H_{\rm int}(t) \simeq - \int d^3x \frac{a^3}{H} A \delta(t-t_0) \dot{\zeta}^{3} ~,
\end{equation}
where $t_0$ is the time at which $|Q_s\Lambda_1|$ reaches its maximum, the constant $A$ carries a mass dimension. The value of $A$ can be evaluated numerically.
Note that $t_0$ lies in the middle of the NEC-violating phase, as evident from Fig. \ref{fig:QsLambdas}. As a result, the one-loop contribution at the end of the second inflationary phase, i.e., $t=t_e$, is
\begin{align}
\label{eq:11term}
    \langle \zeta(\vec p) \zeta(-\vec p) \rangle_{(1,1)} (t_e) \nonumber = & A^2 \frac{a(t_0)^6}{H(t_0)^2} \int \Pi_{i=1}^6 \left[ \frac{d^3k_i}{(2\pi)^3} \right] \delta^{(3)} (\vec{k}_1 + \vec{k}_2 + \vec{k}_3) \delta^{(3)} (\vec{k}_4 + \vec{k}_5 + \vec{k}_6) \\
    & \times \langle \dot{\zeta}_{\vec{k}_1}(t_0) \dot{\zeta}_{\vec{k}_2}(t_0) \dot{\zeta}_{\vec{k}_3}(t_0) \zeta_{\vec{p}} \zeta_{-\vec{p}} \dot{\zeta}_{\vec{k}_1}(t_0) \dot{\zeta}_{\vec{k}_2}(t_0) \dot{\zeta}_{\vec{k}_3}(t_0) \rangle ~.
\end{align}
\begin{align}
\label{eq:02term}
    \langle \zeta(\vec p) \zeta(-\vec p) \rangle_{(0,2)} (t_e)  \nonumber =& A^2 \frac{a(t_0)^6}{H(t_0)^2} \int \Pi_{i=1}^6 \left[ \frac{d^3k_i}{(2\pi)^3} \right] \delta^{(3)} (\vec{k}_1 + \vec{k}_2 + \vec{k}_3) \delta^{(3)} (\vec{k}_4 + \vec{k}_5 + \vec{k}_6) \\
    & \times \langle \zeta_{\vec{p}} \zeta_{-\vec{p}} \dot{\zeta}_{\vec{k}_1}(t_0) \dot{\zeta}_{\vec{k}_2}(t_0) \dot{\zeta}_{\vec{k}_3}(t_0)  \dot{\zeta}_{\vec{k}_1}(t_0) \dot{\zeta}_{\vec{k}_2}(t_0) \dot{\zeta}_{\vec{k}_3}(t_0) \rangle ~.
\end{align}

\subsection{3.3. Dynamics of curvature perturbation and the final one-loop correction}

In the investigation of the one-loop correction to the CMB scale power spectrum, our primary focus lies on the contribution from perturbation modes that exit their horizon around the NEC violation. These perturbation modes are significantly enhanced, thereby contributing to the production of PBHs with masses and abundances of observational interest. In addition, in the calculation of Eqs. (\ref{eq:11term}) and (\ref{eq:02term}), we also need to know the evolution of perturbation modes that were already outside the horizon well before the NEC violation, around $t_0$.

For our purposes, Eq. (\ref{eq:zetace001}) provides a sufficiently good approximation within the time interval of interest to us when calculating $\dot{\zeta}_k(t_0)$ for the perturbation modes that exited the horizon before the second inflationary stage (see Fig. \ref{fig:Comp-zetap002} for an example).
With Eq. (\ref{eq:zetace001}), after evaluating the momentum integration, Eqs. \eqref{eq:11term} and \eqref{eq:02term} become
\begin{align}
    \langle \zeta(\vec p) \zeta(-\vec p) \rangle_{(1,1)} (t_e) & \nonumber = 36 a(t_0)^6 \frac{A^2}{H(t_0)^2} |\zeta_p(t_e)|^2 |\dot{\zeta}_p(t_0)|^2 \int \frac{d^3k}{(2\pi)^3}  |\dot{\zeta}_k(t_0)|^2  |\dot{\zeta}_q(t_0)|^2 \\
    & =  \frac{36A^2}{H(t_0)^2 z_s(t_0)^{12}} |\zeta_p(t_e)|^2 D_p^2 \int \frac{k^2dk}{2\pi^2} D_k^2 D_q^2 ~,
\end{align}
\begin{equation}
    \langle \zeta(\vec p) \zeta(-\vec p) \rangle_{(0,2)} (t_e) = 36 a(t_0)^6 \frac{A^2}{H(t_0)^2} \zeta_p^2(t_e) \dot{\zeta}_p^{\ast 2}(t_0) \int \frac{d^3k}{(2\pi)^3}  |\dot{\zeta}_k(t_0)|^2  |\dot{\zeta}_q(t_0)|^2 ~,
\end{equation}
where the factor $36$ comes from Wick contraction, and $\vec{q} \equiv \vec{k} - \vec{p}$. Since $k \gg p$, we may simply take $q \simeq k$ and $D_q \simeq D_k$. Therefore, we have
\begin{equation}
    |\zeta_p^2(t_e) \dot{\zeta}_p^{\ast 2}(t_0)| = |\zeta_p(t_e)|^2 |\dot{\zeta}_p(t_0)|^2 ~\to~ {\rm Re} [\zeta_p^2(t_e) \dot{\zeta}_p^{\ast 2}(t_0)] \leq |\zeta_p(t_e)|^2 |\dot{\zeta}_p(t_0)|^2 ~.
\end{equation}
As a result, the final one-loop correction \eqref{eq:oneloopexpression} satisfies
\begin{equation}
    \Delta P_{\zeta}(p) \leq \frac{108A^2}{H(t_0)^2 z_s(t_0)^{12}} D_p^2 \int \frac{k^2dk}{2\pi^2} D_k^4 ~.\label{eq:DeltaPInt002}
\end{equation}

\begin{figure}[htbp]
\centering %
\includegraphics[scale=2,width=0.65\textwidth]{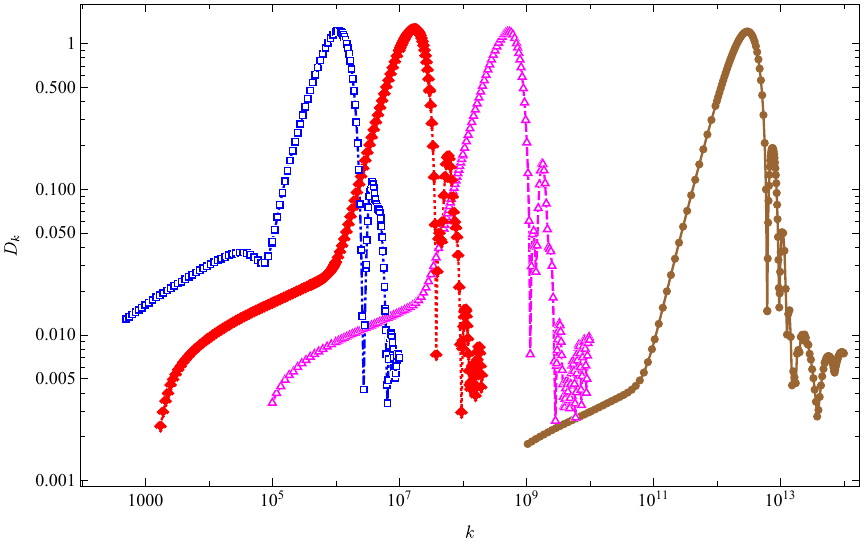}
\caption{The numerical solutions of $D_k$ obtained using the parameter sets corresponding to the blue, red, magenta, and brown curves in Fig. 3 of the main text. We have normalized $D_k$ with a factor $\simeq \text{max}[z_s^{-2}]/\text{max}[\zeta']$ in this plot.}\label{figfDk01}
\end{figure}

Since it is challenging to handle the expression for $D_k$ analytically, we numerically evaluate it and present the result in Fig. \ref{figfDk01}. The points in Fig. \ref{figfDk01} of each color correspond to the curves of the same color in Fig. 3 of the main text.
For each color, we provide $D_k$ within a specific range of $k$. It is important to note that in numerically computing $D_k$, we cannot significantly extend $D_k$ beyond the range of $k$ shown in the plot. This is because the perturbation modes corresponding to larger $k$ exit horizon during the second inflationary stage. For these modes, Eq. (\ref{eq:zetace001}) is no longer applicable around the NEC-violating phase.
Since the power spectrum of perturbation modes with large $k$ is not enhanced, it is reasonable to assume their contribution to the one-loop correction can be neglected. Therefore, when computing the integral in Eq. (\ref{eq:DeltaPInt002}), we are allowed to set a UV cutoff for each color's $D_k$ based on Fig. \ref{figfDk01}.

For the perturbation modes corresponding to smaller $k$ (e.g., $k=p$), Eq. (\ref{eq:zetace001}) holds theoretically. However, since these modes exit the horizon much earlier, their evolving mode (characterized by $D_k$) around the NEC-violating phase is many orders of magnitude smaller than the constant mode, making it numerically infeasible to accurately compute $\dot{\zeta}$ or $\zeta'$, and thus we cannot obtain $D_k$ for smaller $k$ via numerical computation. This is especially true for CMB scale modes where $k=p$.


Fortunately, the analytical expression for the curvature perturbation in the first inflationary stage is known as
\begin{equation}
    \zeta = \frac{\pi H}{2M_{\text{P}}\sqrt{\epsilon k^3}} (1+ik\tau) e^{-ik\tau} ~,\quad ~ \zeta^{\prime} = \frac{\pi H \tau e^{-ik\tau}}{2M_{\text{P}} \sqrt{\epsilon}}\sqrt{k} ~.
\end{equation}
The curvature perturbation $\zeta$ and its time derivative $\zeta'$ should be continuous, allowing us to impose the matching condition around the transition (denoted as $\tau = \tau_1$) from the first inflationary stage to the NEC-violating stage for $\zeta^{\prime}$, i.e.,
\begin{equation}
    |\zeta'(\tau_1)| \simeq \frac{D_k}{z_s^2(\tau_1)} ~\Rightarrow~ D_k \simeq  \frac{\pi H(\tau_1) \tau_1}{2M_{\text{P}} \sqrt{\epsilon(\tau_1) }} z_s^2(\tau_1) \sqrt{k} \propto k^{\frac{1}{2}} ~.
\end{equation}
Since we neglected the details of the transition, the matching method may lose some accuracy. As shown by the red and magenta curves in Fig. \ref{figfDk01}, $D_k$ may decrease even faster than $k^{1/2}$ as $k$ decreases.
However, in the subsequent estimation of one-loop corrections, we still choose to estimate the value of $D_p$ by interpolating the result in Fig. \ref{figfDk01} using the relation $D_p \propto p^{1/2}$. We should keep in mind that such an estimation may lead to an overestimation of the one-loop corrections.


Taking $p=10^{-2}$ $\text{Mpc}^{-1}$ as an example, we obtain $D_p=5.56\times 10^{-9}$, $1.08\times10^{-6}$, $4.14\times10^{-6}$, and $5.75\times10^{-5}$ for the brown, magenta, red, and blue curves in Fig. \ref{figfDk01}, respectively.
Using Eq. (\ref{eq:DeltaPInt002}), we find the one-loop corrections to the CMB scale power spectrum approximately satisfy $\Delta P_{\zeta}(p)<4.57\times10^{-3}$, $1.36\times10^{-2}$, $2.86\times10^{-2}$ and $2.91$ for the brown, magenta, red, and blue curves in Fig. 3 of the main text, respectively.

\subsection{3.4. Discussions}

It appears that for a peak value of $P_\zeta\simeq 10^{-2}$, the enhancement of the power spectrum at larger scales may result in a smaller one-loop correction to the CMB scale power spectrum.
The discussion in Kristiano$\&$Yokoyama \cite{Kristiano:2022maq} regarding one-loop corrections to the CMB scale power spectrum implies that the blue curve should be excluded, while the remaining three curves can survive.
It is worth noting that there is a suspicion of overestimation in the one-loop correction results for the blue and brown curves. This is because the blue and brown $D_k$ curves in Fig. \ref{figfDk01} were truncated due to numerical precision issues before an upward convex trend (i.e., $d^2D_k/dk^2<0$ and $dD_k/d k>0$) appeared.

Additionally, it is important to emphasize that when presenting the power spectrum curves in Fig. 3 of the main text, none of the parameters were deliberately adjusted to yield smaller one-loop corrections to the CMB scale power spectrum. Under such a circumstance, the upper limits of the one-loop corrections to the power spectrum at the CMB scale provided by our scenario have already reached or even fallen below the order of $10^{-2}$. Therefore, within a certain parameter space, our scenario may have the potential to circumvent the ``No-Go'' proposed for single-field USR models in \cite{Kristiano:2022maq}.

In the context of computing the one-loop corrections to the power spectrum at the CMB scale, there exists a notable distinction between our scenario and USR. In the USR scenario, the coefficient of the dominant term in the cubic action, i.e., $\dot{\eta}$, and $\zeta$ almost simultaneously reach their maximum values at the end of USR. In contrast, in our scenario, when the coefficient functions $Q_s|\Lambda_i|$ in the cubic action reach their maximum values, $\zeta$ or its derivatives are far from reaching their maximum values, as exhibited in Fig. \ref{fig:QsLambdas}. When $\zeta$ or its derivatives reach their maximum values, those coefficient functions have already decreased to sufficiently small values.
It is this characteristic of our scenario that may provide a potentially novel approach to yielding smaller one-loop corrections to the CMB scale power spectrum.

In our aforementioned calculations, we have made some simplifications, such as using simplified parameterizations and only computing the dominant term in the cubic action.
Therefore, despite the differences with the USR scenario, obtaining conclusive proof that the result in \cite{Kristiano:2022maq} does not invalidate our scenario necessitates further investigation in the future.

\end{document}